\documentclass[12pt,a4paper]{article}

\usepackage[latin1]{inputenc}
\usepackage{lineno}  % for line numbering during review
\usepackage{graphicx}  % to include figures (can also use other packages)
\usepackage{xspace} % To avoid problems with missing or double spaces after
\usepackage{booktabs} % To be able to define nicer tables.
\usepackage{multirow}
\usepackage{color}
\usepackage{colortbl}
\usepackage{amsmath} % Adds a large collection of math symbols
\usepackage{ifthen} % for conditional statements
\usepackage{subfigure}
\usepackage{lscape} % for large tables

\usepackage{indentfirst}

\newboolean{pdflatex}
\setboolean{pdflatex}{true} % use this if using non-eps figures
\newboolean{articletitles}
\setboolean{articletitles}{true} % False removes titles in references

\newboolean{uprightparticles}
\setboolean{uprightparticles}{false} %Set to true to get roman particle symbols

\usepackage{amssymb}
\usepackage{amsfonts}
\usepackage{upgreek} % Adds in support for greek letters in roman typeset

\usepackage{fink}
\usepackage{xifthen}
\usepackage{ifpdf}

\usepackage{hyperref}    % Hyperlinks in references
\usepackage[all]{hypcap} % Internal hyperlinks to floats.

\def\ux85 {UX85\xspace}

\ifthenelse{\boolean{uprightparticles}}%
{

 \def\PDelta      {\ensuremath{\Delta}\xspace}                 
 \def\PXi      {\ensuremath{\Xi}\xspace}                 
 \def\PLambda      {\ensuremath{\Lambda}\xspace}                 
 \def\PSigma      {\ensuremath{\Sigma}\xspace}                 
 \def\POmega      {\ensuremath{\Omega}\xspace}                 
 \def\PUpsilon      {\ensuremath{\Upsilon}\xspace}                 
 
 \def\PB      {\ensuremath{\mathrm{B}}\xspace}                 
                  
 \def\PD      {\ensuremath{\mathrm{D}}\xspace}

 \def\PK      {\ensuremath{\mathrm{K}}\xspace}

 \def\Pi      {\ensuremath{\mathrm{i}}\xspace}

}
{

 \mathchardef\PDelta="7101
 \mathchardef\PXi="7104
 \mathchardef\PLambda="7103
 \mathchardef\PSigma="7106
 \mathchardef\POmega="710A
 \mathchardef\PUpsilon="7107
                  
 \def\PB      {\ensuremath{B}\xspace}                 
                  
 \def\PD      {\ensuremath{D}\xspace}

 \def\PK      {\ensuremath{K}\xspace}

 \def\Pi      {\ensuremath{i}\xspace}

}

\def\kaon  {\ensuremath{\PK}\xspace}
  \def\Kbar  {\kern 0.2em\overline{\kern -0.2em \PK}{}\xspace}

\def\Kz    {\ensuremath{\kaon^0}\xspace}
\def\Kzb   {\ensuremath{\Kbar^0}\xspace}
\def\KzKzb {\ensuremath{\Kz \kern -0.16em \Kzb}\xspace}
\def\Kp    {\ensuremath{\kaon^+}\xspace}
\def\Km    {\ensuremath{\kaon^-}\xspace}

\def\KpKm  {\ensuremath{\Kp \kern -0.16em \Km}\xspace}

  \def\Dbar    {\kern 0.2em\overline{\kern -0.2em \PD}{}\xspace}
\def\D       {\ensuremath{\PD}\xspace}

\def\Dz      {\ensuremath{\D^0}\xspace}
\def\Dzb     {\ensuremath{\Dbar^0}\xspace}
\def\DzDzb   {\ensuremath{\Dz {\kern -0.16em \Dzb}}\xspace}
\def\Dp      {\ensuremath{\D^+}\xspace}
\def\Dm      {\ensuremath{\D^-}\xspace}

\def\DpDm    {\ensuremath{\Dp {\kern -0.16em \Dm}}\xspace}

  \def\Bbar    {\kern 0.18em\overline{\kern -0.18em \PB}{}\xspace}

  \def\Y#1S{\ensuremath{\PUpsilon{(#1S)}}\xspace}% no space before {...}!

\def\to                 {\ensuremath{\rightarrow}\xspace}

\def\CP                {\ensuremath{C\!P}\xspace}

\def\AT#1     {\ensuremath{A_{\mathrm{T}}^{#1}}\xspace}           % 2

\def\C#1      {\ensuremath{\mathcal{C}_{#1}}\xspace}                       % 9
\def\Cp#1     {\ensuremath{\mathcal{C}_{#1}^{'}}\xspace}                    % 7
\def\Ceff#1   {\ensuremath{\mathcal{C}_{#1}^{\mathrm{(eff)}}}\xspace}        % 9  
\def\Cpeff#1  {\ensuremath{\mathcal{C}_{#1}^{'\mathrm{(eff)}}}\xspace}       % 7
\def\Ope#1    {\ensuremath{\mathcal{O}_{#1}}\xspace}                       % 2
\def\Opep#1   {\ensuremath{\mathcal{O}_{#1}^{'}}\xspace}                    % 7

\newcommand{\tev}{\ensuremath{\mathrm{\,Te\kern -0.1em V}}\xspace}
\newcommand{\gev}{\ensuremath{\mathrm{\,Ge\kern -0.1em V}}\xspace}
\newcommand{\mev}{\ensuremath{\mathrm{\,Me\kern -0.1em V}}\xspace}
\newcommand{\kev}{\ensuremath{\mathrm{\,ke\kern -0.1em V}}\xspace}
\newcommand{\ev}{\ensuremath{\mathrm{\,e\kern -0.1em V}}\xspace}
\newcommand{\gevc}{\ensuremath{{\mathrm{\,Ge\kern -0.1em V\!/}c}}\xspace}
\newcommand{\mevc}{\ensuremath{{\mathrm{\,Me\kern -0.1em V\!/}c}}\xspace}
\newcommand{\gevcc}{\ensuremath{{\mathrm{\,Ge\kern -0.1em V\!/}c^2}}\xspace}
\newcommand{\gevgevcccc}{\ensuremath{{\mathrm{\,Ge\kern -0.1em V^2\!/}c^4}}\xspace}
\newcommand{\mevcc}{\ensuremath{{\mathrm{\,Me\kern -0.1em V\!/}c^2}}\xspace}

\def\gsim{{~\raise.15em\hbox{$>$}\kern-.85em
          \lower.35em\hbox{$\sim$}~}\xspace}
\def\lsim{{~\raise.15em\hbox{$<$}\kern-.85em
          \lower.35em\hbox{$\sim$}~}\xspace}

\def\tell1  {TELL1\xspace}
\def\ukl1   {UKL1\xspace}

\usepackage{mciteplus}

\newcolumntype{e}{@{ $\pm$ }l}         % Column that contains \pm.
\newcolumntype{R}{>{$}r<{$}}           % Right  aligned math column.
\newcolumntype{L}{>{$}l<{$}}           % Left   aligned math column.
\newcolumntype{C}{>{$}c<{$}}           % Center aligned math column.
\newcolumntype{E}{@{ $\pm$ }>{$}l<{$}} % Error left aligned math column.

\newcommand\uncheckbox{\makebox[0pt][l]{$\square$}\raisebox{.15ex}{\hspace{0.1em}\hphantom{$\checkmark$}}}

\usepackage{wasysym}     

\newcommand{\plot}[2][width=.95\textwidth]{
  \ifpdf
  \IfFileExists{\finkbase/png/#2.png}
  {
    \includegraphics[#1,type=png,ext=.png,read=.png]{\finkbase/png/#2}
  }
  {
    \IfFileExists{\finkbase/jpg/#2.jpg}
    {
      \includegraphics[#1,type=jpg,ext=.jpg,read=.jpg]{\finkbase/jpg/#2}
    }
    {
      \includegraphics[#1,type=png,ext=.png,read=.png]{format/blank}
    }
  }
  \else
  \IfFileExists{\finkbase/eps/#2.eps}{
    \includegraphics[#1,type=eps,ext=.eps,read=.eps]{\finkbase/eps/#2}
  }{
    \includegraphics[#1,type=eps,ext=.eps,read=.eps]{format/blank}
  }
  \fi
}

\definecolor{orange}{rgb}{1,0.5,0}

\newcommand{\re}[2][()] {\ifthenelse{\equal{#1}{()}}{{\ensuremath{{\rm \, Re}}\left(#2\right)}}
                                                    {{\ensuremath{{\rm \, Re}}\left[#2\right]}}}
\newcommand{\im}[2][()] {\ifthenelse{\equal{#1}{()}}{{\ensuremath{{\rm \, Im}}\left(#2\right)}}
                                                    {{\ensuremath{{\rm \, Im}}\left[#2\right]}}}
\newcommand{\gauss}[3]  {
                        \ifthenelse{\isempty{#3}}
                        {G\left( #1; \mu^{#1,#2}     , \sigma^{#1,#2}      \right)}
                        {G\left( #1; \mu^{#1,#2}_{#3}, \sigma^{#1,#2}_{#3} \right)}
                        }
\newcommand{\bgauss}[3] {
                        \ifthenelse{\isempty{#3}}
                        {G_b\left( #1; \mu^{#1,#2}     , \sigma_l^{#1,#2}     , \sigma_r^{#1,#2}      \right)}
                        {G_b\left( #1; \mu^{#1,#2}_{#3}, \sigma_l^{#1,#2}_{#3}, \sigma_r^{#1,#2}_{#3} \right)}
                        }
\newcommand{\gaussC}[4] {
                        \ifthenelse{\isempty{#4}}
                        {G_c\left( #1,#2;\mu^{#1,#3}     ,\sigma^{#1,#3}     ,
                        \mu^{#2,#3}     ,\sigma^{#2,#3}     ,\kappa     \right)}
                        {G_c\left( #1,#2;\mu^{#1,#3}_{#4},\sigma^{#1,#3}_{#4},
                        \mu^{#2,#3}_{#4},\sigma^{#2,#3}_{#4},\kappa_{#4}\right)}
                        }
\newcommand{\johnson}[3]{
                        \ifthenelse{\isempty{#3}}
                        {J_{S_U}\left(#1;\mu^{#1,#2}     ,\sigma^{#1,#2}     ,
                        \gamma^{#1,#2}     ,\delta^{#1,#2}     \right)}
                        {J_{S_U}\left(#1;\mu^{#1,#2}_{#3},\sigma^{#1,#2}_{#3},
                        \gamma^{#1,#2}_{#3},\delta^{#1,#2}_{#3}\right)}
                        }
\newcommand{\argus}[3]  {
                        \ifthenelse{\isempty{#3}}
                        {B\left(#1;m_\pi,\xi^{#1,#2}     \right)}
                        {B\left(#1;m_\pi,\xi^{#1,#2}_{#3}\right)}
                        }
\newcommand{\mitj}[1]   {
                        \ifthenelse{\isempty{#1}}
                        {\langle \mu      \rangle}
                        {\langle \mu^{#1} \rangle}
                        }
\newcommand{\desv}[1]   {
                        \ifthenelse{\isempty{#1}}
                        {\langle \sigma      \rangle}
                        {\langle \sigma^{#1} \rangle}
                        }
\newcommand{\kst}[3]    {
                        \ifthenelse{\isempty{#2}}
                        {$K^{\star #1}     (#3)$\xspace}
                        {$K^{\star #1}_{#2}(#3)$\xspace}
                        }

\newcommand{\mzket}   {\left| \vphantom{\bar{M}^0}     {M}^0 \right\rangle}
\newcommand{\mzbket}  {\left|                      \bar{M}^0 \right\rangle}

\newcommand{\mpmcpket}{\left|     {M}_\pm^{\mathrm{CP}}      \right\rangle}
\newcommand{\mzkett}  {\left|     {M}^0(t)                   \right\rangle}
\newcommand{\mzbkett} {\left| \bar{M}^0(t)                   \right\rangle}

\newcommand{\dzket}   {\left|       \vphantom{\bar{D}^0}     {D}^0      \right\rangle}
\newcommand{\dzbket}  {\left|                            \bar{D}^0      \right\rangle}
\newcommand{\dzkett}  {\left|       \vphantom{\bar{D}^0}     {D}^0(t)   \right\rangle}
\newcommand{\dzbkett} {\left|                            \bar{D}^0(t)   \right\rangle}
\newcommand{\dzkettd} {\left|       \vphantom{\bar{D}^0}     {D}^0(t_D) \right\rangle}
\newcommand{\dzbkettd}{\left|                            \bar{D}^0(t_D) \right\rangle}
\newcommand{\dzbra}   {\left\langle \vphantom{\bar{D}^0}     {D}^0      \right|}
\newcommand{\dzbbra}  {\left\langle                      \bar{D}^0      \right|}
\newcommand{\dpmcpket}{\left|     {D}_\pm^{\mathrm{CP}}      \right\rangle}

\newcommand{\dpket}   {\left|           {D}_+                      \right\rangle}
\newcommand{\dmket}   {\left|           {D}_-                      \right\rangle}
\newcommand{\dpmket}  {\left|           {D}_\pm                    \right\rangle}
\newcommand{\bpmket}  {\left|           {B}_\pm                    \right\rangle}

\newcommand{\kzket}   {\left|       \vphantom{\bar{K}^0}     {K}^0      \right\rangle}
\newcommand{\kzbket}  {\left|                            \bar{K}^0      \right\rangle}

\newcommand{\kzkettk} {\left|       \vphantom{\bar{K}^0}     {K}^0(t_K) \right\rangle}
\newcommand{\kzbkettk}{\left|                            \bar{K}^0(t_K) \right\rangle}
\newcommand{\kzbra}   {\left\langle \vphantom{\bar{K}^0}     {K}^0      \right|}
\newcommand{\kzbbra}  {\left\langle                      \bar{K}^0      \right|}

\newcommand{\fket}    {\left|       f                        \right\rangle}
\newcommand{\fbra}    {\left\langle f                        \right|}

\newcommand{\mpmket}  {\left|     {M}_\pm                    \right\rangle}
\newcommand{\mpmkett} {\left|     {M}_\pm(t)                 \right\rangle}

\newcommand{\phimixt}{\phi_{\mathrm{mix}}(t)}

\newcommand{\gmix}   {G_{\mathrm{mix}}}

\newcommand{\gpmt}   {G_{\pm}(t)}

\newcommand{\eixicp} {e^{ i\xi_{\mathrm{CP}}}}

\newcommand{\qoverp}{\frac{q}{p}}
\newcommand{\poverq}{\frac{p}{q}}

\begin{document}

\begin{titlepage}

\vspace*{-1.5cm}

\hspace*{-0.5cm}

\vspace*{4.0cm}

{\bf\boldmath\huge
\begin{center}
The hyperbolic rotation group of neutral meson mixing and \CP violation
\end{center}
}

\vspace*{2.0cm}

\begin{center}
Jordi Garra Tico$^1$
\bigskip\\
{\it\footnotesize
$ ^1$Cavendish Laboratory, University of Cambridge, Cambridge, United Kingdom\\
}
\end{center}

\vspace{\fill}

\begin{abstract}
  Neutral meson mixing and \CP violation are very well known weak processes that involve
  decays to meson states that are, in general, a superposition of flavor eigenstates. This
  paper describes a mathematical interpretation of the time-dependent mixing amplitudes as
  a complex hyperbolic rotation of the time evolution of those amplitudes without mixing,
  which involves a Lie group $SO(1,1,\mathbb{C})$.

  This allows a geometric interpretation of mixing as a curve into the $SO(1,1,\mathbb{C})$
  manifold, parameterized with the proper decay time, where \CP violation is the image of
  this curve at $t = 0$.

  To show the power of this new interpretation, it is applied to several aspects of the measurement
  of the CKM angle $\gamma$ in $B$ decays to neutral $D$ mesons. On one hand, the charm mixing
  correction on the $CPV$ parameters is derived. On the other hand, it is shown how the
  expressions used in GLW, ADS and GGSZ methods are affected by charm mixing. Finally, the
  complete example with both charm and strange mixing and $CPV$ is described.

\end{abstract}

\vspace*{2.0cm}
\vspace{\fill}

\end{titlepage}

\thispagestyle{empty}

\newpage
\setcounter{page}{2}
\mbox{~}

\cleardoublepage

\section{Introduction}

Mixing occurs in weak decays of neutral mesons with flavor eigenstates that are not self-conjugate.
All the information about mixing is encoded in a complex mixing coefficient $z$, that is defined in
terms of the Hamiltonian eigenvalues, and characterizes the time evolution of any neutral meson
eigenstates.

\CP violation may also have an effect on the time evolution of any neutral meson eigenstates, either
through different amplitudes to the same final \CP eigenstate (direct) or through a difference
between the Hamiltonian and \CP eigenstates (indirect).

This paper describes a parameterization that can be used to describe mixing and \CP violation in a
consistent and simplified way. In particular, it describes an interpretation that exploits the fact
that the neutral $D$ mesons produced in $B$ meson decays are a superposition of flavor eigenstates
that can be used to extract the CKM matrix phase $\gamma$, and these states undergo mixing as they
evolve in time.

\section{The mixing formalism} \label{sc.mixing}

\subsection{States and conventions}
There are three bases of states that are relevant to the discussion on mixing and \CP violation
of neutral mesons $M$.
\begin{itemize}
  \item The flavor eigenstates $\mzket$ and $\mzbket$ have a definite flavor content. The \CP operator
    acts on these states with an arbitrary phase,
    \begin{align}
      CP \mzket  &= e^{+i\xi_{\mathrm{CP}}} \mzbket, \\
      CP \mzbket &= e^{-i\xi_{\mathrm{CP}}} \mzket.
    \end{align}
    The most common conventions are $e^{i \xi_{\mathrm{CP}}} = \pm 1$.
  \item The \CP eigenstates $\mpmcpket$, such that $CP \mpmcpket = \pm \mpmcpket$. They
    can be expressed in terms of the flavor eigenstates as
    \begin{equation}
      \mpmcpket = \frac{\mzket \pm e^{i\xi_{\mathrm{CP}}} \mzbket}{\sqrt{2}}.
    \end{equation}
  \item The Hamiltonian eigenstates $\mpmket$. They can be expressed in terms of the flavor eigenstates as
    \begin{equation}
      \mpmket = p \mzket \pm s \, q \mzbket, \label{eq.heigenstates}
    \end{equation}
    which defines the change of base parameters $p$ and $q$, where $s$ is an arbitrary sign. These states are
    properly normalized if $\left| p \right|^2 + \left| q \right|^2 = 1$. If \CP is preserved, then
    $\mpmket = \mpmcpket$, and
    \begin{equation}
      \qoverp = s \, \eixicp.
    \end{equation}
    In this paper, the $\mpmket$ states are intentionally noted to represent the Hamiltonian eigenstates
    that are closest to the $\mpmcpket$ \CP eigenstates, respectively.
\end{itemize}

For the rest of this paper, the convention chosen is $\eixicp = 1$ and $s = +1$.

\subsection{Mixing equations}

The time evolution of the Hamiltonian eigenstates is given by the Schrödinger equation,
\begin{equation}
  i \frac{d}{dt} \mpmket = \hat{H} \mpmket,
\end{equation}
where $\hat{H}$ is the effective Hamiltonian operator
\begin{equation}
  \hat{H} = \hat{M} - i \frac{\hat{\Gamma}}{2}.
\end{equation}
The solution of the time evolution can be expressed, therefore, as
\begin{equation}
  \mpmkett = \theta(t) e^{-i H_\pm t} \mpmket = h_\pm(t) \mpmket,
\end{equation}
where $H_\pm$ are the Hamiltonian eigenvalues for $\mpmket$. This defines the $h_\pm(t)$ functions.
For convenience, it is useful to define
\begin{align}
  m      &= \frac{m_- + m_+}{2},                          \\
  \Gamma &= \frac{\Gamma_- + \Gamma_+}{2},                \\
  H_0    &= \frac{H_- + H_+}{2} = m - i \frac{\Gamma}{2}.
\end{align}
With these definitions it is useful to express
\begin{equation} \label{eq.hpmt}
  h_\pm(t) = \theta(t) e^{-i H_\pm t} = \theta(t) e^{-i H_0 t} e^{\pm z \frac{\Gamma t}{2}} = \theta(t) e^{-i m t} e^{-\frac{\Gamma t}{2}} e^{\pm \phimixt},
\end{equation}
where
\begin{equation}
  z = \frac{i \left( H_- - H_+ \right)}{\Gamma} = \frac{\Gamma_- - \Gamma_+}{2\Gamma} + i \frac{m_- - m_+}{\Gamma},
\end{equation}
and
\begin{equation}
  \phimixt = \frac{z \Gamma t}{2}.
\end{equation}

The $z$ complex coefficient contains all the necessary information to describe mixing. In the charm sector, equations
are typically expressed in terms of the mixing parameters $x_D$ and $y_D$, such that
\begin{equation}
  z = - ( y_D + i x_D ).
\end{equation}
In the bottom sector, the parameters of interest are typically $\Delta m$, defined as the mass difference between the
heavy and the light eigenstates, $\Delta m = m_h - m_l$, and $\Delta \Gamma$, defined as the width difference between
the light and the heavy eigenstates, $\Delta \Gamma = \Gamma_l - \Gamma_h$. With this convention, $\Delta m$ is
positively defined and
\begin{equation}
  z = - \frac{\Delta \Gamma}{2\Gamma} + i \frac{\Delta m}{\Gamma}.
\end{equation}
In this paper, the real and imaginary parts of $z$ are noted $x$ and $y$, respectively, i.e. $z = x + i\,y$.

The time dependence of all the mixing observables can be expressed in terms of $\left| h_\pm(t) \right|^2$ and $h_+ h_-^\star$.
Therefore, the term $e^{-i m t}$ in equation (\ref{eq.hpmt}) vanishes in all the observables and can be omitted,
\begin{equation}
  h_\pm(t) = \theta(t) e^{-\frac{\Gamma t}{2}} e^{\pm \phimixt}.
\end{equation}

The time evolution of the flavor eigenstates can be derived from their expression in terms of Hamiltonian eigenstates
(\ref{eq.heigenstates}),
\begin{align}
  \mzkett  &= \frac{h_+(t) + h_-(t)}{2} \mzket  + \qoverp \frac{h_+(t) - h_-(t)}{2} \mzbket, \label{eq.mzt}  \\
  \mzbkett &= \frac{h_+(t) + h_-(t)}{2} \mzbket + \poverq \frac{h_+(t) - h_-(t)}{2} \mzket.  \label{eq.mzbt}
\end{align}
For convenience, the $g_\pm(t)$ functions are defined as
\begin{equation}
  g_\pm(t) = \frac{h_+(t) \pm h_-(t)}{2}
                                          = \theta(t) \, e^{-\frac{\Gamma t}{2}} ~^{\cosh}_{\sinh}\left[ \phimixt \right],
\end{equation}
so equations (\ref{eq.mzt}) and (\ref{eq.mzbt}) can be written as
\begin{align}
  p \mzkett  &= g_+(t) \, p \mzket  + g_-(t) \, q \mzbket, \\
  q \mzbkett &= g_+(t) \, q \mzbket + g_-(t) \, p \mzket,
\end{align}
\newcommand{\coshzgth}{\cosh \phimixt}
\newcommand{\sinhzgth}{\sinh \phimixt}
which can be expressed in matrix form as
\begin{equation}
  \begin{pmatrix}
    p \mzkett  \vphantom{\eixicp q \bar{A}(t) e^{\frac{\Gamma t}{2}}} \\
    q \mzbkett \vphantom{\eixicp q \bar{A}(t) e^{\frac{\Gamma t}{2}}}
  \end{pmatrix}
  =
  \begin{pmatrix}
    \vphantom{\eixicp p \mzkett  e^{\frac{\Gamma t}{2}}} \coshzgth & \sinhzgth \\
    \vphantom{\eixicp q \mzbkett e^{\frac{\Gamma t}{2}}} \sinhzgth & \coshzgth
  \end{pmatrix}
  \begin{pmatrix}
    e^{-\frac{\Gamma t}{2}} \, p \mzket   \\
    e^{-\frac{\Gamma t}{2}} \, q \mzbket
  \end{pmatrix}.
\end{equation}

This allows us to interpret mixing as a complex hyperbolic rotation of the time evolution of the
states vector without mixing, given by
\begin{equation}
  e^{-\frac{\Gamma t}{2}}
  \begin{pmatrix}
    p \mzket   \\
    q \mzbket
  \end{pmatrix}.
\end{equation}

The time evolution of the neutral mesons can, therefore, be expressed as
\begin{equation} \label{eq.qgq}
  \begin{pmatrix}
    \mzkett  \vphantom{\eixicp q \bar{A}(t) e^{\frac{\Gamma t}{2}}} \\
    \mzbkett \vphantom{\eixicp q \bar{A}(t) e^{\frac{\Gamma t}{2}}}
  \end{pmatrix}
  =
  e^{-\frac{\Gamma t}{2}}
  Q^{-1}
  \gmix
  Q
  \begin{pmatrix}
    \mzket   \\
    \mzbket
  \end{pmatrix},
\end{equation}
where
\begin{equation}
  \gmix =
  G[ \phimixt ] =
  \begin{pmatrix}
    \vphantom{\eixicp p \bar{A}(t) e^{\frac{\Gamma t}{2}}} \coshzgth & \sinhzgth \\
    \vphantom{\eixicp q \bar{A}(t) e^{\frac{\Gamma t}{2}}} \sinhzgth & \coshzgth
  \end{pmatrix},
  \qquad
  Q =
  \begin{pmatrix}
    p & 0 \\
    0 & q
  \end{pmatrix}.
\end{equation}

The $g_\pm(t)$ functions have the following useful properties:
\begin{align}
  \left| g_\pm(t) \right|^2 &= \theta(t) \, e^{- \Gamma t} \, \frac{ \cosh{(x \Gamma t)} \pm \cos{(y \Gamma t)} }{2}, \\
  g_+^\star(t) g_-(t)       &= \theta(t) \, e^{- \Gamma t} \, \frac{ \sinh{(x \Gamma t)} + i \sin{(y \Gamma t)} }{2}.
\end{align}

For the rest of the paper, the $\theta(t)$ term is not written explicitly and the expressions that follow are only
valid for positive $t$. For negative $t$, all amplitudes and states are zero.

\newcommand{\ampdz} {\left\langle f \! \left| \vphantom{\bar{M}^0} H \right| \!     {M}^0 \right\rangle}
\newcommand{\ampdzb}{\left\langle f \! \left| \vphantom{\bar{M}^0} H \right| \! \bar{M}^0 \right\rangle}

\newcommand{\atanh}{\mathrm{atanh}}

The amplitudes of the $\mzket$ and $\mzbket$ decays into a final state $\fket$,
$\ampdz$ and $\ampdzb$, are given by
\begin{equation} \label{eq.transform}
  \begin{pmatrix}
    p     {A}(t) \vphantom{\eixicp q \bar{A}(t) e^{\frac{\Gamma t}{2}}} \\
    q \bar{A}(t) \vphantom{\eixicp q \bar{A}(t) e^{\frac{\Gamma t}{2}}}
  \end{pmatrix}
  =
  \gmix(t)
  \begin{pmatrix}
    p     {A} \, e^{-\frac{\Gamma t}{2}} \\
    q \bar{A} \, e^{-\frac{\Gamma t}{2}}
  \end{pmatrix},
\end{equation}
which can be interpreted as a complex hyperbolic rotation of the amplitude vector time evolution without mixing,
\begin{equation}
  \begin{pmatrix}
    p     {A} \, e^{-\frac{\Gamma t}{2}} \\
    q \bar{A} \, e^{-\frac{\Gamma t}{2}}
  \end{pmatrix}.
\end{equation}
The hyperbolic rotation $\gmix$ is an element of the Lie group $SO(1,1,\mathbb{C})$, which is also a
differentiable manifold. The time variable $t$ parameterizes a \textbf{curve} into the manifold.
This gives mixing a geometric interpretation. In section \S\ref{sc.cpveq} it is shown that this
interpretation can also be extended to $CPV$.

Because the Lorentz group with one space dimension is $SO^+(1,1,\mathbb{R})$, the mixing group
has properties similar to those of special relativity. In particular, one can write the time evolution of
\begin{equation}
  \lambda = \qoverp \frac{\bar{A}}
                         {    {A}}
\end{equation}
as
\begin{equation}
  \lambda(t) = \qoverp \frac{\bar{A}(t)}
                            {    {A}(t)}
             = \frac{\lambda + \tanh\phimixt}
                    {1 + \lambda \tanh\phimixt}
\end{equation}
which, not surprisingly, has the same form as the formula of addition of velocities in special relativity.
In analogy to the concept of rapidity, one can define the hyperbolic magnitude
\begin{equation}
  \phi_\lambda = \atanh\left(\lambda\right),
\end{equation}
whose time evolution is just additive,
\begin{equation} \label{eq.philambdat}
  \phi_\lambda(t) = \phi_\lambda + \phimixt.
\end{equation}

In this hyperbolic space, the time evolution of $\phi_\lambda$ is a point moving from its initial value
towards a constant direction determined by $z$ and $\Gamma$ and, since $\phimixt$ is a linear function
in time, this move takes place at a constant pace.

Equation (\ref{eq.transform}) imposes a mixing invariant, given by Minkowski squaring the transformed vector,
\begin{equation}
  p^2 A^2(t) - q^2 \bar{A}^2(t) = \left( p^2 A^2 - q^2 \bar{A}^2 \right) e^{-\Gamma t}.
\end{equation}

\section{CPV equations}
\label{sc.cpveq}

In experimental measurements of the CKM angle $\gamma$, it is important to derive the time dependence
of amplitudes in the whole decay chain. To illustrate the formalism described above, it will be
useful to go through the complete example of a $B$ meson decay to a neutral $D$ meson.

Measurements of the CKM angle $\gamma$ involve decays where a neutral $D$ meson is produced as a superposition
of flavor eigenstates $\dzket$ and $\dzbket$,
\begin{align}
  \dmket &\propto \dzket  + z_- \dzbket, \\
  \dpket &\propto \dzbket + z_+ \dzket,
\end{align}
where $\dmket$ denotes a state dominated by the $\dzket$ component and $\dpket$ a state dominated by
the $\dzbket$ component,
\begin{equation}
  \left| \frac{\left\langle                      \bar{D}^0 \left|\vphantom{\bar{D}^0}\right. \!D_-\right\rangle}
              {\left\langle \vphantom{\bar{D}^0}     {D}^0 \left|\vphantom{\bar{D}^0}\right. \!D_-\right\rangle} \right| < 1,
  \qquad
  \left| \frac{\left\langle \vphantom{\bar{D}^0}     {D}^0 \left|\vphantom{\bar{D}^0}\right. \!D_+\right\rangle}
              {\left\langle                      \bar{D}^0 \left|\vphantom{\bar{D}^0}\right. \!D_+\right\rangle} \right| < 1.
\end{equation}
Decays such as $B^\pm \rightarrow D K^\pm$ produce $\dpmket$ states, respectively, but the sign in $\dpmket$ may not
correspond to the charge of the $B$ mother particle, such as in $B^0 \rightarrow D K^{*0}$ decays. For this reason, to
avoid confusion between the $B$ mother charge and the flavor content of its daughters, the initial state is denoted
$\bpmket$, instead of $\left| B^\pm \right\rangle$.

The time evolution of the product of amplitudes from the initial $\bpmket$ state to a final state $\fket$ is given by
\begin{equation} \label{eq.abpmt}
  A^{B_\pm}_f(t) = \fbra
  \mathcal{H}
  \begin{pmatrix}
    \dzkett & \dzbkett
  \end{pmatrix}
  \begin{pmatrix}
    \vphantom{\frac{1^1}{1^1}} \dzbra \\ \vphantom{\frac{1^1}{1^1}} \dzbbra
  \end{pmatrix}
  \mathcal{H}
  \bpmket
\end{equation}

The time evolution of the neutral $D$ component in this equation is given by equation (\ref{eq.qgq}), so
\begin{equation}
  \begin{aligned}
    \fbra \mathcal{H}
    \begin{pmatrix}
      \dzkett & \dzbkett
    \end{pmatrix}
    &=
    e^{-\frac{\Gamma t}{2}}
    \fbra \mathcal{H}
    \begin{pmatrix}
      \dzket & \dzbket
    \end{pmatrix}
    Q \gmix Q^{-1} \\
    &=
    e^{-\frac{\Gamma t}{2}}
    \begin{pmatrix}
      A^D_f & A^{\bar{D}}_f
    \end{pmatrix}
    Q \gmix Q^{-1},
  \end{aligned}
\end{equation}
which uses the fact that $G^\intercal = G$ and $Q^\intercal = Q$. On the other hand,
\begin{equation}
  \begin{pmatrix}
    \vphantom{\frac{1^1}{1^1}} \dzbra \\ \vphantom{\frac{1^1}{1^1}} \dzbbra
  \end{pmatrix}
  \mathcal{H}
  \bpmket
  =
  A^\pm_{\tilde{D}}
  \begin{pmatrix}
    1     & z_\pm \\
    z_\pm & 1
  \end{pmatrix}
  \begin{pmatrix}
    \delta_{\pm -} \\
    \delta_{\pm +}
  \end{pmatrix},
\end{equation}
where $\delta_{ab}$ is the Kronecker delta, the vector on the right hand side is used to select
the appropriate column of the matrix, and $A^\pm_{\tilde{D}}$ is used to represent either of
\begin{equation}
  A^-_D         = \left\langle     {D}^0 | \mathcal{H} | B^- \right\rangle,
  \qquad
  A^+_{\bar{D}} = \left\langle \bar{D}^0 | \mathcal{H} | B^+ \right\rangle.
\end{equation}

By defining
\begin{align}
  \beta_\pm  &= \left( \qoverp \right)^{\!\!\pm 1} z_\pm, \\
  \gamma_\pm &= \frac{1}{\displaystyle \sqrt{1 - \beta_\pm^2}}, \\
  \phi_\pm   &= \atanh{\beta_\pm},
\end{align}
it is easy to check that
\begin{equation}
  Q^{-1}
  \begin{pmatrix}
    1     & z_\pm \\
    z_\pm & 1
  \end{pmatrix}
  \begin{pmatrix}
    \delta_{\pm -} \\
    \delta_{\pm +}
  \end{pmatrix}
  =
  \gamma_\pm^{-1}
  G_\pm
  Q^{-1}
  \begin{pmatrix}
    \delta_{\pm -} \\
    \delta_{\pm +}
  \end{pmatrix},
\end{equation}
where $G_\pm = G( \phi_\pm )$.

With all these definitions, equation (\ref{eq.abpmt}) becomes
\begin{equation} \label{eq.apm}
  A^{B_\pm}_f(t) =
  e^{-\frac{\Gamma t}{2}}
  \frac{A^\pm_{\tilde{D}}}
       {\gamma_\pm}
  \begin{pmatrix}
    A^D_f & A^{\bar{D}}_f
  \end{pmatrix}
  Q
  \gpmt
  Q^{-1}
  \begin{pmatrix}
    \delta_{\pm -} \\
    \delta_{\pm +}
  \end{pmatrix},
\end{equation}
with
\begin{equation}
  \gpmt = \gmix G_\pm = G[ \phi_\pm(t) ],
\end{equation}
where, analogously to equation (\ref{eq.philambdat}), the time evolution of the $\phi_\pm$ observables is given by
\begin{equation} \label{eq.phipmt}
  \phi_\pm(t) = \phi_\pm + \phimixt.
\end{equation}

In some cases it may be useful to even include the terms of the amplitude vector into the hyperbolic rotation matrix,
\begin{equation}
  A^{B_\pm}_f(t) =
  e^{-\frac{\Gamma t}{2}}
  \frac{A^\pm_{\tilde{D}}}
       {\gamma_\pm}
  \frac{A^{\tilde{D}}_f}
       {\gamma_\pm^\lambda}
  \begin{pmatrix}
    \delta_{\pm -} & \delta_{\pm +}
  \end{pmatrix}
  Q
  G_{\lambda\pm}(t)
  Q^{-1}
  \begin{pmatrix}
    \delta_{\pm -} \\
    \delta_{\pm +}
  \end{pmatrix},
\end{equation}
where the function used for $G_{\lambda\pm}(t)$ is given by
\begin{equation}
  \phi^{\lambda_D}_\pm(t) = \phi_\pm + \phi_\pm^{\lambda_D} + \phimixt,
\end{equation}
and
\begin{equation}
  \phi_\pm^{\lambda_D} = \atanh( \lambda_D^{\mp 1} ),
\end{equation}
and
\begin{equation} \label{eq.lambdaD}
  \lambda_D = \qoverp \frac{A^{\bar{D}}_f}{A^{D}_f}.
\end{equation}
However, in other cases $A^{D}_f$ may be zero or very small in some regions of the phase space, and $\lambda_D$
in equation (\ref{eq.lambdaD}) may not be well defined or be numerically unstable in these regions. To preserve
generality as much as possible, the following sections do not include the terms of the amplitude vector in the
hyperbolic rotation matrix.

\subsection{Charm mixing correction on $CPV$ parameters}

Equations (\ref{eq.apm}) allow to express the $A_\pm(t)$ amplitudes using a common formalism with
mixing, as described in \S\ref{sc.mixing}. In order to take all experimental effects into account, a simultaneous
fit for the mixing and \CP violation parameters is advised. However, considering the large effort it may represent,
a mixing correction of the \CP violation parameters is easily deduced from (\ref{eq.phipmt}).
Denoting as $\phi_\pm^{\mathrm{m}}$ the measured hyperbolic \CP violation parameters in a fit that ignores mixing,
\begin{equation}
  \phi_\pm^{\mathrm{m}} = \left\langle \phi_\pm(t) \right\rangle = \phi_\pm + \left\langle \phimixt \right\rangle,
\end{equation}
so the actual hyperbolic \CP violation parameters can be obtained as a correction on the measured ones as
\begin{equation}
  \phi_\pm
           = \phi_\pm^{\mathrm{m}} - \frac{z \left\langle \Gamma t \right\rangle}{2}.
\end{equation}

Notice that, if $\phi_\pm$ parameters are used, the only information necessary to apply a mixing correction to a
result is a measurement of $\left\langle \Gamma t \right\rangle$, which is valid for \textbf{any} lifetime distribution.
If $z_\pm$ was used instead, one would have to calculate
\begin{equation}
  \beta_\pm = \left\langle \frac{\beta_\pm^{\mathrm{m}} - \tanh\left[\phi_{\mathrm{mix}}(t)\right]}
                                { 1 - \beta_\pm^{\mathrm{m}} \tanh\left[\phi_{\mathrm{mix}}(t)\right]} \right\rangle,
\end{equation}
and, in this case, a measurement of $\left\langle \Gamma t \right\rangle$ would not be enough to calculate the correction.

\subsection{Specific equations for established methodologies}

Charm mixing has effects on the GLW, ADS and GGSZ observables. This subsection describes
how the equations relevant to these methods are modified in the presence of charm mixing.

The time-dependent measurement of $\gamma$ in $B_s \to D_s^\mp K^\pm$ depends on $B_s$
mixing and it can also be easily expressed using the formalism in \S\ref{sc.cpveq}.

\subsubsection{Common trace} \label{sc.commontrace}

Since the most common amplitudes are proportional to
\begin{equation}
\gamma_\pm^{-1}
\begin{pmatrix}
  a & b
\end{pmatrix}
Q \gpmt Q^{-1}
\begin{pmatrix}
  \delta_{\pm-} \\ \delta_{\pm+}
\end{pmatrix},
\end{equation}
for some values of $a$ and $b$, it is useful to calculate
\begin{equation}
  \mathrm{Tr}
  \left[
    e^{-\Gamma t}
    \left|\gamma_\pm\right|^{-2}
    \begin{pmatrix}
      \delta_{\pm-} & \delta_{\pm+}
    \end{pmatrix}
    Q^{\star -1} \gpmt^\star Q^\star
    \begin{pmatrix}
      a^\star \\ b^\star
    \end{pmatrix}
    \begin{pmatrix}
      a & b
    \end{pmatrix}
    Q \gpmt Q^{-1}
    \begin{pmatrix}
      \delta_{\pm-} \\ \delta_{\pm+}
    \end{pmatrix}
  \right].
\end{equation}
The result can be expressed as
\begin{equation}
  \mathrm{Tr}\left[ \dots \right]
  =
  \delta_{\pm-} \left| a \, u_-(t) + b \, \zeta_-(t) \right|^2 +
  \delta_{\pm+} \left| b \, u_+(t) + a \, \zeta_+(t) \right|^2,
\end{equation}
where
\begin{align}
  u_\pm(t)     &= e^{-\frac{\Gamma t}{2}}                                      \frac{\cosh\phi_\pm(t)}{\cosh\phi_\pm}, \\
  \zeta_\pm(t) &= e^{-\frac{\Gamma t}{2}} \left(\frac{q}{p}\right)^{\!\!\mp 1} \frac{\sinh\phi_\pm(t)}{\cosh\phi_\pm},
\end{align}
which satisfy
\begin{align}
  u_\pm(0)     &= 1, \\
  \zeta_\pm(0) &= z_\pm.
\end{align}

For a later section, it is also useful to define
\begin{equation}
  \xi_\pm(t) = e^{-\frac{\Gamma t}{2}} \frac{\sinh\phi_\pm(t)}{\cosh\phi_\pm}.
\end{equation}

\subsubsection{GLW equations}

The GLW method \cite{glw1,glw2} uses, on one hand, states that are accessible from only one of the flavor eigenstates,
either $\dzket$ or $\dzbket$, such that
\begin{align}
  A^{    {D}}_{    {f}} &= \left\langle f_{    {D}} | \mathcal{H} |     {D}^0 \right\rangle,     \\
  A^{\bar{D}}_{\bar{f}} &= \left\langle f_{\bar{D}} | \mathcal{H} | \bar{D}^0 \right\rangle,     \\
  A^{    {D}}_{\bar{f}} &= \left\langle f_{\bar{D}} | \mathcal{H} |     {D}^0 \right\rangle = 0, \\
  A^{\bar{D}}_{    {f}} &= \left\langle f_{    {D}} | \mathcal{H} | \bar{D}^0 \right\rangle = 0,
\end{align}
and, on the other hand, states that are accessible from either of the $CP$ eigenstates $\dpmcpket$, such that
\begin{align}
  A^{\mathrm{CP}}_\pm =& \left\langle f_\pm | \mathcal{H} | D^\mathrm{CP}_\pm \right\rangle,     \\
                       & \left\langle f_\mp | \mathcal{H} | D^\mathrm{CP}_\pm \right\rangle = 0.
\end{align}

If the initial $D$ meson state is a superposition $\dpmket$, then
\begin{align}
  \left\langle f_{    {D}} | \mathcal{H} | D_-(t) \right\rangle
  &=
  e^{-\frac{\Gamma t}{2}}
  A^{    {D}}_{    {f}} \gamma_-^{-1}
  \begin{pmatrix}
    1 & 0
  \end{pmatrix}
  Q_D^{-1} G_-(t) Q_D
  \begin{pmatrix}
    1 \\ 0
  \end{pmatrix}, \\
  \left\langle f_{\bar{D}} | \mathcal{H} | D_+(t) \right\rangle
  &=
  e^{-\frac{\Gamma t}{2}}
  A^{\bar{D}}_{\bar{f}} \gamma_+^{-1}
  \begin{pmatrix}
    0 & 1
  \end{pmatrix}
  Q_D^{-1} G_+(t) Q_D
  \begin{pmatrix}
    0 \\ 1
  \end{pmatrix}, \\
  \left\langle f_\pm | \mathcal{H} | D_-(t) \right\rangle
  &=
  e^{-\frac{\Gamma t}{2}}
  A^{\mathrm{CP}}_\pm \gamma_-^{-1}
  \begin{pmatrix}
    1 & 0
  \end{pmatrix}
  Q_D^{-1} G_-(t) Q_D
  \begin{pmatrix}
    \frac{1}{\sqrt{2}} \\ \pm\frac{1}{\sqrt{2}}
  \end{pmatrix}, \\
  \left\langle f_\pm | \mathcal{H} | D_+(t) \right\rangle
  &=
  e^{-\frac{\Gamma t}{2}}
  A^{\mathrm{CP}}_\pm \gamma_+^{-1}
  \begin{pmatrix}
    0 & 1
  \end{pmatrix}
  Q_D^{-1} G_+(t) Q_D
  \begin{pmatrix}
    \frac{1}{\sqrt{2}} \\ \pm\frac{1}{\sqrt{2}}
  \end{pmatrix}.
\end{align}

The observables of interest are
\begin{align}
  R_{\mathrm{CP}}^\pm
  &=
  \frac{\Gamma\left( B^- \rightarrow D_\pm^\mathrm{CP} h^- \right) +
        \Gamma\left( B^+ \rightarrow D_\pm^\mathrm{CP} h^+ \right)}
       {\Gamma\left( B^- \rightarrow     {D}^0         h^- \right) +
        \Gamma\left( B^+ \rightarrow \bar{D}^0         h^+ \right)} \\
  &=
  \frac{\left|\left\langle f_\pm       | \mathcal{H} | D_-(t) \right\rangle\right|^2 +
        \left|\left\langle f_\pm       | \mathcal{H} | D_+(t) \right\rangle\right|^2}
       {\left|\left\langle f_{    {D}} | \mathcal{H} | D_-(t) \right\rangle\right|^2 +
        \left|\left\langle f_{\bar{D}} | \mathcal{H} | D_+(t) \right\rangle\right|^2} \\
  &=
  \frac{1}{2} \left| \frac{A^{\mathrm{CP}}_\pm}{A^D_f} \right|^2
  \frac{\left|u_-(t) \pm \zeta_-(t)\right|^2 + \left|u_+(t) \pm \zeta_+(t)\right|^2}
       {\left|u_-(t)               \right|^2 + \left|u_+(t)               \right|^2},
\end{align}
and
\begin{align}
  A_{\mathrm{CP}}^\pm
  &=
  \frac{\Gamma\left( B^- \rightarrow D_\pm^\mathrm{CP} h^- \right) -
        \Gamma\left( B^+ \rightarrow D_\pm^\mathrm{CP} h^+ \right)}
       {\Gamma\left( B^- \rightarrow D_\pm^\mathrm{CP} h^- \right) +
        \Gamma\left( B^+ \rightarrow D_\pm^\mathrm{CP} h^+ \right)} \\
  &=
  \frac{\left|\left\langle f_\pm | \mathcal{H} | D_-(t) \right\rangle\right|^2 -
        \left|\left\langle f_\pm | \mathcal{H} | D_+(t) \right\rangle\right|^2}
       {\left|\left\langle f_\pm | \mathcal{H} | D_-(t) \right\rangle\right|^2 +
        \left|\left\langle f_\pm | \mathcal{H} | D_+(t) \right\rangle\right|^2} \\
  &=
  \frac{\left|u_-(t) \pm \zeta_-(t)\right|^2 - \left|u_+(t) \pm \zeta_+(t)\right|^2}
       {\left|u_-(t) \pm \zeta_-(t)\right|^2 + \left|u_+(t) \pm \zeta_+(t)\right|^2}.
\end{align}

Notice that, at $t=0$,
\begin{align}
  \frac{\left|u_-(t) \pm \zeta_-(t)\right|^2 + \left|u_+(t) \pm \zeta_+(t)\right|^2}
       {\left|u_-(t)               \right|^2 + \left|u_+(t)               \right|^2}
  =
  \frac{\left|1 \pm z_- \right|^2 + \left|1 \pm z_+\right|^2}{2}
  &=
  1 + r^2 \pm 2 r \cos\delta \cos\gamma, \\
  \frac{\left|u_-(t) \pm \zeta_-(t)\right|^2 - \left|u_+(t) \pm \zeta_+(t)\right|^2}
       {\left|u_-(t) \pm \zeta_-(t)\right|^2 + \left|u_+(t) \pm \zeta_+(t)\right|^2}
  =
  \frac{\left|1 \pm z_- \right|^2 - \left|1 \pm z_+\right|^2}
       {\left|1 \pm z_- \right|^2 + \left|1 \pm z_+\right|^2}
  &=
  \frac{\pm 2 r \sin\delta \sin\gamma}{1 + r^2 \pm 2 r \cos\delta \cos\gamma}.
\end{align}

\subsubsection{ADS equations}

The ADS method \cite{ads3} uses states that are accessible from both flavor eigenstates.
With the convention $CP \dzket = \dzbket$ and assuming no direct $CP$ violation in the $D$ decay,
one can define the $\rho$ ratio of amplitudes as
\begin{equation}
  \rho
  =
  \frac{A^{\bar{D}}_{    {f}}}
       {A^{    {D}}_{    {f}}}
  =
  \frac{A^{    {D}}_{\bar{f}}}
       {A^{\bar{D}}_{\bar{f}}}.
\end{equation}

If the initial $D$ meson state is a superposition $\dpmket$, then
\begin{align}
  \left\langle f_{    {D}} | \mathcal{H} | D_-(t) \right\rangle
  &=
  e^{-\frac{\Gamma t}{2}}
  A^{    {D}}_{    {f}} \gamma_-^{-1}
  \begin{pmatrix}
    1 & 0
  \end{pmatrix}
  Q_D^{-1} G_-(t) Q_D
  \begin{pmatrix}
    1 \\ \rho
  \end{pmatrix}, \\
  \left\langle f_{\bar{D}} | \mathcal{H} | D_-(t) \right\rangle
  &=
  e^{-\frac{\Gamma t}{2}}
  A^{\bar{D}}_{\bar{f}} \gamma_-^{-1}
  \begin{pmatrix}
    1 & 0
  \end{pmatrix}
  Q_D^{-1} G_-(t) Q_D
  \begin{pmatrix}
    \rho \\ 1
  \end{pmatrix}, \\
  \left\langle f_{    {D}} | \mathcal{H} | D_+(t) \right\rangle
  &=
  e^{-\frac{\Gamma t}{2}}
  A^{    {D}}_{    {f}} \gamma_+^{-1}
  \begin{pmatrix}
    0 & 1
  \end{pmatrix}
  Q_D^{-1} G_+(t) Q_D
  \begin{pmatrix}
    1 \\ \rho
  \end{pmatrix}, \\
  \left\langle f_{\bar{D}} | \mathcal{H} | D_+(t) \right\rangle
  &=
  e^{-\frac{\Gamma t}{2}}
  A^{\bar{D}}_{\bar{f}} \gamma_+^{-1}
  \begin{pmatrix}
    0 & 1
  \end{pmatrix}
  Q_D^{-1} G_+(t) Q_D
  \begin{pmatrix}
    \rho \\ 1
  \end{pmatrix}
\end{align}

One can define
\begin{alignat}{2}
  \Gamma^\pm_\mathrm{fav}
  &=
  \Gamma\left( B^\pm \rightarrow D_\mathrm{fav} \, h^\pm \right)
  &&=
  \left\{
  \begin{aligned}
    \left|
    \left\langle f_{    {D}} | \mathcal{H} | D_-(t) \right\rangle
    \right|^2
    && \text{for $D_-$}
    \\
    \left|
    \left\langle f_{\bar{D}} | \mathcal{H} | D_+(t) \right\rangle
    \right|^2
    && \text{for $D_+$}
  \end{aligned}
  \right.
  \\
  &&&=
  \left| A^\pm_{\tilde{D}} A^D_f \right|^2
  \left|        u_\pm(t) + \rho \, \zeta_\pm(t)\right|^2
  ,
  \\
  \Gamma^\pm_\mathrm{sup}
  &=
  \Gamma\left( B^\pm \rightarrow D_\mathrm{sup} \, h^\pm \right)
  &&=
  \left\{
  \begin{aligned}
    \left|
    \left\langle f_{\bar{D}} | \mathcal{H} | D_-(t) \right\rangle
    \right|^2
    && \text{for $D_-$}
    \\
    \left|
    \left\langle f_{    {D}} | \mathcal{H} | D_+(t) \right\rangle
    \right|^2
    && \text{for $D_+$}
  \end{aligned}
  \right.
  \\
  &&&=
  \left| A^\pm_{\tilde{D}} A^D_f \right|^2
  \left|\rho \, u_\pm(t) +         \zeta_\pm(t)\right|^2
  ,
\end{alignat}
where $D_\mathrm{sup}$ and $D_\mathrm{fav}$ refer to suppresed and favored decay modes of
the produced $D$ meson.

The observables of interest are
\begin{equation}
  R_{\mathrm{ADS}}^\pm
  =
  \frac{\Gamma^\pm_\mathrm{sup}}
       {\Gamma^\pm_\mathrm{fav}}
  =
  \left|
    \frac{\rho \, u_\pm(t) +         \zeta_\pm(t)}
         {        u_\pm(t) + \rho \, \zeta_\pm(t)}
  \right|^2.
\end{equation}

It is also frequent to use the observables
\begin{alignat}{2}
  R_{\mathrm{ADS}}
  &=
  \frac{\Gamma^-_\mathrm{sup} + \Gamma^+_\mathrm{sup}}
       {\Gamma^-_\mathrm{fav} + \Gamma^+_\mathrm{fav}}
  &&=
  \frac{\left|\rho \, u_-(t) +         \zeta_-(t)\right|^2 +
        \left|\rho \, u_+(t) +         \zeta_+(t)\right|^2}
       {\left|        u_-(t) + \rho \, \zeta_-(t)\right|^2 +
        \left|        u_+(t) + \rho \, \zeta_+(t)\right|^2},
  \\
  A_{\mathrm{ADS}}^\mathrm{sup}
  &=
  \frac{\Gamma^-_\mathrm{sup} - \Gamma^+_\mathrm{sup}}
       {\Gamma^-_\mathrm{sup} + \Gamma^+_\mathrm{sup}}
  &&=
  \frac{\left|\rho \, u_-(t) +         \zeta_-(t)\right|^2 -
        \left|\rho \, u_+(t) +         \zeta_+(t)\right|^2}
       {\left|\rho \, u_-(t) +         \zeta_-(t)\right|^2 +
        \left|\rho \, u_+(t) +         \zeta_+(t)\right|^2},
  \\
  A_{\mathrm{ADS}}^\mathrm{fav}
  &=
  \frac{\Gamma^-_\mathrm{fav} - \Gamma^+_\mathrm{fav}}
       {\Gamma^-_\mathrm{fav} + \Gamma^+_\mathrm{fav}}
  &&=
  \frac{\left|        u_-(t) + \rho \, \zeta_-(t)\right|^2 -
        \left|        u_+(t) + \rho \, \zeta_+(t)\right|^2}
       {\left|        u_-(t) + \rho \, \zeta_-(t)\right|^2 +
        \left|        u_+(t) + \rho \, \zeta_+(t)\right|^2}.
\end{alignat}

\subsubsection{GGSZ equations}

The GGSZ method \cite{ggsz} uses $3$ or more body decays to final states that can be accessed from any of
$\dzket$ or $\dzbket$.
In this case, there may be regions of the phase space where one of the amplitudes $A^D_f$ or
$A^{\bar{D}}_f$ is zero or very small. Because of this, it may be inconvenient to use terms
where one of the amplitudes is in the denominator and the expression for the common trace
obtained in \S\ref{sc.commontrace} may be undefined or numerically unstable.

The probability distribution function $p_\pm(t)$ is proportional to $\left| A^{B_\pm}_f(t) \right|^2$,
\begin{equation}
  \left| A^{B_\pm}_f(t) \right|^2
  =
  e^{-\Gamma t}
  \left|
    \frac{A^\pm_{\tilde{D}}}
         {\gamma_\pm}
  \right|^2
  \begin{pmatrix}
    \delta_{\pm -} &
    \delta_{\pm +}
  \end{pmatrix}
  Q^{-1\star}
  \gpmt^\star
  Q^\star
  \begin{pmatrix}
    {A^D_f}^\star \\ {A^{\bar{D}}_f}^\star
  \end{pmatrix}
  \begin{pmatrix}
    A^D_f & A^{\bar{D}}_f
  \end{pmatrix}
  Q
  \gpmt
  Q^{-1}
  \begin{pmatrix}
    \delta_{\pm -} \\
    \delta_{\pm +}
  \end{pmatrix}.
\end{equation}

To simplify this expression, it is convenient to define
\begin{equation}
  \delta_\pm
  =
  \begin{pmatrix}
    \delta_{\pm -} \\
    \delta_{\pm +}
  \end{pmatrix}
  \begin{pmatrix}
    \delta_{\pm -} &
    \delta_{\pm +}
  \end{pmatrix}
  =
  \begin{pmatrix}
    \delta_{\pm-} & 0             \\
    0             & \delta_{\pm+}
  \end{pmatrix},
\end{equation}
and
\begin{equation}
  \begin{pmatrix}
    A_+ \\
    A_-
  \end{pmatrix}
  =
  \frac{1}{2}
  S Q
  \begin{pmatrix}
    A^D_f \\
    A^{\bar{D}}_f
  \end{pmatrix},
\end{equation}
where
\begin{equation}
  S =
  \begin{pmatrix}
    1 &  1 \\
    1 & -1
  \end{pmatrix},
\end{equation}
such that
\begin{equation}
  Q
  \begin{pmatrix}
    A^D_f \\
    A^{\bar{D}}_f
  \end{pmatrix}
  =
  S
  \begin{pmatrix}
    A_+ \\
    A_-
  \end{pmatrix}.
\end{equation}
By defining
\begin{equation}
  M =
  \begin{pmatrix}
    \left| A_+ \right|^2 & A_+^\star A_- \\
    A_+ A_-^\star        & \left| A_- \right|^2
  \end{pmatrix}
\end{equation}
and
\begin{equation}
  \Psi_\pm(t)
  =
  e^{-\Gamma t} S \gpmt \delta_\pm \gpmt^\star S
  =
  e^{-\Gamma t}
  \begin{pmatrix}
        e^{  2   \re[]{\phi_\pm(t)}} & \mp e^{2 i \im[]{\phi_\pm(t)}} \\
    \mp e^{- 2 i \im[]{\phi_\pm(t)}} &     e^{- 2 \re[]{\phi_\pm(t)}}
  \end{pmatrix}
  =
  \begin{pmatrix}
    \psi_+(t)             & \mp \psi_i(t) \\
    \mp {\psi_i}^\star(t) & \psi_-(t)
  \end{pmatrix},
\end{equation}
one can see that
\begin{equation}
  \left| A^{B_\pm}_f(t) \right|^2
  =
  \left|
    \frac{A^\pm_{\tilde{D}}}
         {\gamma_\pm}
  \right|^2
  \frac{1}{\left| p \, \delta_{\pm-} + q \, \delta_{\pm+} \right|^2}
  \mathrm{Tr} \left[ \Psi_\pm(t) M \right].
\end{equation}

The magnitudes involved in the previous equation include
\begin{align}
  A_\pm       &= \frac{p A^D_f \pm q A^{\bar{D}}_f}{2}, \\
  \psi_\pm(t) &= e^{\pm 2 \re{\phi_\pm}}e^{-\left( 1 \mp x \right) \Gamma t},  \\
  \psi_i  (t) &= e^{2 i   \im{\phi_\pm}}e^{-\left( 1 - i y \right) \Gamma t}.
\end{align}
To account for lifetime resolution and acceptance effects, the $\psi_\pm(t)$ and $\psi_i(t)$ functions should
be replaced by the result of convolving them by the lifetime resolution function and multiplying the result
by the lifetime acceptance function, as described in \cite{eqsmixing}.

The normalized time-dependent amplitude pdf is given by
\begin{equation}
  p_\pm(t)
  =
  \frac{\mathrm{Tr} \left[ \Psi_\pm(t) M   \right]}
       {\mathrm{Tr} \left[ N_\pm^t     N^M \right]},
\end{equation}
where
\begin{equation}
  N_\pm^t = \int \mathrm{d}t \, \Psi_\pm(t),
\end{equation}
and
\begin{equation}
  N^M = \int \mathrm{d\mathcal{P}} \, M.
\end{equation}

\subsubsection{Time dependent equations}

\newcommand{\sizeof}[2]{\lefteqn{#2}\phantom{#1}}

The time dependent method \cite{td} uses $B_s$ decays that undergo mixing before they decay, such as
$B_s \to D_s^\mp K^\pm$. The amplitudes of the $B_s$ decay are given by
\begin{align}
  A^{B_s}_{D_s^\pm}(t)
  &= A^{B_s}_{D_s^\pm} \left[ g_+(t) + \sizeof{\lambda_\pm^{-1}}{\lambda_\pm} g_-(t) \right],
  \\
  A^{\bar{B}_s}_{D_s^\pm}(t)
  &= A^{\bar{B}_s}_{D_s^\pm} \left[ g_+(t) + \lambda_\pm^{-1} g_-(t) \right],
\end{align}
where $t$ represents the $B_s$ lifetime.
These expressions can also be written as
\begin{equation}
  \begin{pmatrix}
    p A^{B_s}_{D_s^\pm}(t) \\
    q A^{\bar{B}_s}_{D_s^\pm}(t)
  \end{pmatrix}
  =
  e^{-\frac{\Gamma t}{2}}
  \frac{p A^{B_s}_{D_s^\pm}}{\gamma_\lambda}
  \begin{pmatrix}
    \cosh\phi^{B_s}_\pm(t) \\
    \sinh\phi^{B_s}_\pm(t)
  \end{pmatrix}
  =
  p A^{B_s}_{D_s^\pm}
  \begin{pmatrix}
    u  ^{B_s}_\pm(t) \\
    \xi^{B_s}_\pm(t)
  \end{pmatrix},
\end{equation}
where $\Gamma$ represents the $B_s$ decay width and
\begin{align}
  \phi^{B_s}_\pm(t)  &= \phi_{\lambda_\pm} + \phi_{\mathrm{mix}}^{B_s}(t), \\
  \phi_{\lambda_\pm} &= \atanh\left( \lambda_\pm \right), \\
  \lambda_\pm                                          &= \frac{q}{p} e^{-i \phi_s} z_\pm^{\mp1}.
\end{align}

\subsection{A complete example with charm and strange mixing and $CPV$}

This example describes all charm and strange mixing and $CPV$ effects in decays of the form
$B \to D h$, where $D \to K h^+ h^-$ and $K \to \pi^+ \pi^-$. The vast majority of neutral kaons
that decay to $\pi^+ \pi^-$ are $K_s$, but because neither $K_s$ or $K_l$ are pure $CP$ eigenstates,
there may be a small amount of $K_l$ that decay to $\pi^+ \pi^-$ as well, which results in
\begin{equation}
  \eta = \frac{\left\langle\pi\pi \left| \mathcal{H} \right| K_l \right\rangle}
              {\left\langle\pi\pi \left| \mathcal{H} \right| K_s \right\rangle}
\end{equation}
being non-zero.

In measurements of $CP$ violation in $B$ decays, kaon mixing and $CP$ violation are negligible
in practice and, as of 2017, it is impossible to take these effects into account in $\gamma$
measurements because all the information available on the $D$ decay amplitude has been obtained
assuming no $CP$ violation in the kaon. This is, therefore, a purely academic exercise.

In this example, $t_D$ is the lifetime of the $D$ meson and $t_K$ is the lifetime of the $K$ meson.

{\footnotesize
\begin{align}
  A^{B_\pm}_f(t_D, t_K)
  &=
  \left\langle \pi\pi \right|
  \mathcal{H}
  \begin{pmatrix}
    \kzkettk & \kzbkettk
  \end{pmatrix}
  \begin{pmatrix}
    \vphantom{\frac{1^1}{1^1}} \!\kzbra \\ \vphantom{\frac{1^1}{1^1}} \!\kzbbra
  \end{pmatrix}
  \mathcal{H}
  \begin{pmatrix}
    \dzkettd & \dzbkettd
  \end{pmatrix}
  \begin{pmatrix}
    \vphantom{\frac{1^1}{1^1}} \!\dzbra \\ \vphantom{\frac{1^1}{1^1}} \!\dzbbra
  \end{pmatrix}
  \mathcal{H}
  \bpmket \\
  &
  \begin{aligned}
    \;\!=& \; e^{-\frac{\Gamma_D t_D}{2}}
    e^{-\frac{\Gamma_K t_K}{2}}
    \\&
    \left\langle \pi\pi \right|
    \mathcal{H}
    \begin{pmatrix}
      \kzket & \!\!\!\kzbket
    \end{pmatrix}
    Q_K G_K^\mathrm{mix} Q_K^{-1}
    \begin{pmatrix}
      \vphantom{\frac{1^1}{1^1}} \!\kzbra \\ \vphantom{\frac{1^1}{1^1}} \!\kzbbra
    \end{pmatrix}
    \mathcal{H}
    \begin{pmatrix}
      \dzket & \!\!\!\dzbket
    \end{pmatrix}
    Q_D G_D^\mathrm{mix} Q_D^{-1}
    \begin{pmatrix}
      \vphantom{\frac{1^1}{1^1}} \!\dzbra \\ \vphantom{\frac{1^1}{1^1}} \!\dzbbra
    \end{pmatrix}
    \mathcal{H}
    \bpmket
   \end{aligned}
  \\
  &=
  e^{-\frac{\Gamma_D t_D}{2}}
  e^{-\frac{\Gamma_K t_K}{2}}
  \frac{A^\pm_{\tilde{D}}}
       {\gamma_\pm}
  \left\langle \pi\pi \right|
  \mathcal{H}
  \begin{pmatrix}
    \kzket & \!\!\!\kzbket
  \end{pmatrix}
  Q_K G_K^\mathrm{mix} Q_K^{-1}
  A_0
  Q_D G_{\pm}(t_D) Q_D^{-1}
  \begin{pmatrix}
    \delta_{\pm-} \\
    \delta_{\pm+}
  \end{pmatrix} \\
  &=
  e^{-\frac{\Gamma_D t_D}{2}}
  e^{-\frac{\Gamma_K t_K}{2}}
  \frac{A^\pm_{\tilde{D}} A^K_{\pi\pi}}
       {\gamma_\pm}
  \begin{pmatrix}
    1 & \rho
  \end{pmatrix}
  Q_K G_K^\mathrm{mix} Q_K^{-1}
  A_0
  Q_D G_{\pm}(t_D) Q_D^{-1}
  \begin{pmatrix}
    \delta_{\pm-} \\
    \delta_{\pm+}
  \end{pmatrix} \\
  &=
  e^{-\frac{\Gamma_D t_D}{2}}
  e^{-\frac{\Gamma_K t_K}{2}}
  \frac{A^\pm_{\tilde{D}} A^K_{\pi\pi}}
       {\gamma_\pm \gamma_\lambda}
  \begin{pmatrix}
    1 & 0
  \end{pmatrix}
  Q_K G_\lambda G_K^\mathrm{mix} Q_K^{-1}
  A_0
  Q_D G_{\pm}(t_D) Q_D^{-1}
  \begin{pmatrix}
    \delta_{\pm-} \\
    \delta_{\pm+}
  \end{pmatrix} \\
  &=
  e^{-\frac{\Gamma_D t_D}{2}}
  e^{-\frac{\Gamma_K t_K}{2}}
  \frac{A^\pm_{\tilde{D}} A^K_{\pi\pi}}
       {\gamma_\pm \gamma_\lambda}
  \begin{pmatrix}
    1 & 0
  \end{pmatrix}
  Q_K G_{\lambda}(t_K) Q_K^{-1}
  A_0
  Q_D G_{\pm}(t_D) Q_D^{-1}
  \begin{pmatrix}
    \delta_{\pm-} \\
    \delta_{\pm+}
  \end{pmatrix},
\end{align}
}
where
\begin{align}
  A_0 &=
  \begin{pmatrix}
    \vphantom{\frac{1^1}{1^1}} \kzbra \\ \vphantom{\frac{1^1}{1^1}} \kzbbra
  \end{pmatrix}
  \mathcal{H}
  \begin{pmatrix}
    \dzket & \dzbket
  \end{pmatrix}
  =
  \begin{pmatrix}
    A^D_{    {K}} & A^{\bar{D}}_{    {K}} \\
    A^D_{\bar{K}} & A^{\bar{D}}_{\bar{K}}
  \end{pmatrix}, \\
  \rho           &= \frac{A^{\bar{K}}_{\pi\pi}}
                         {A^{    {K}}_{\pi\pi}}, \\
  \lambda        &= \left( \frac{q}{p} \right)_{\!\!K} \rho = \frac{ 1 - \eta }{ 1 + \eta }, \\
  \gamma_\lambda &= \frac{1}{ \sqrt{ 1 - \lambda^2 }}, \\
  \phi_\lambda   &= \atanh{\lambda}, \\
  \phi_\lambda(t_K) &= \phi_\lambda + \phi^K_\mathrm{mix}(t_K), \\
  G_\lambda      &= G\left[ \phi_\lambda \right],
  \\
  G_\lambda(t_K) &= G\left[ \phi_\lambda(t_K) \right] = G_\lambda \, G_K^\mathrm{mix}(t_K).
\end{align}

The squared amplitude becomes
\begin{equation}
  \left| A^{B_\pm}_f(t_D, t_K) \right|^2 =
  \left|
    \frac{A^\pm_{\tilde{D}} A^K_{\pi\pi}}
         {\gamma_\pm \gamma_\lambda}
  \right|^2
  \left|
    \frac{p_K}
         {p_D \delta_{\pm-} + q_D \delta_{\pm+}}
  \right|^2
  \mathrm{Tr}\left[ \Psi_\lambda^\star( t_K ) A \Psi_\pm( t_D ) A^\dag \right],
\end{equation}
where
\begin{align}
  \Psi_\lambda( t_K ) &=
  \begin{pmatrix}
    \psi_+^K( t_K )          & \psi_i^K( t_K ) \\
    \psi_i^{K\star}( t_K ) & \psi_-^K( t_K )
  \end{pmatrix}, \\
  A &= \frac{S}{2} Q_k^{-1} A_0 Q_D \frac{S}{2}.
\end{align}

In this case, it is convenient to define the hypermatrix
\begin{equation}
  M_{ijkl} = A^{\star}_{ij} A_{kl},
\end{equation}
such that
\begin{equation}
  \mathrm{Tr}\left[ \Psi_\lambda^\star( t_K ) A \Psi_\pm( t_D ) A^\dag \right]
  =
  \Psi_\pm^{ij}( t_D ) \Psi_\lambda^{kl}( t_K ) M_{ljki}.
\end{equation}

With this hypermatrix, the normalized pdf becomes
\begin{equation}
  p_\pm( t_D, t_K )
  =
  \frac{\Psi_\pm^{ij}( t_D ) \Psi_\lambda^{kl}( t_K ) M_{ljki}}
       {N_\pm^{ij} N_\lambda^{kl} N^M_{ljki}},
\end{equation}
where
\begin{align}
  N_\pm^{ij}     &= \int \mathrm{d}t \, \Psi_\pm^{ij}(t),     \\
  N_\lambda^{ij} &= \int \mathrm{d}t \, \Psi_\lambda^{ij}(t), \\
  N^M_{ijkl}     &= \int \mathrm{d\mathcal{P}} \, M_{ijkl}.
\end{align}

\section{Conclusions}

The mixing expressions have been written in terms of new hyperbolic parameters that allow for an
interpretation of the time evolution of decay amplitudes as a complex hyperbolic rotation of a
vector of flavor amplitudes of decays without mixing. The definition of hyperbolic variables, in
analogy with rapidities in special relativity, simplifies the composition of mixing and \CP
violating effects in a succession of particle decays.

These hyperbolic expressions have been used to obtain the charm mixing correction to the
\CP-violating parameters $\phi_\pm$ or $z_\pm$ involved in several measurements of the CKM angle $\gamma$.
To properly account for mixing and \CP violating effects in the background, or experimental resolution
and acceptance effects, a simultaneous fit of charm mixing and \CP-violating parameters $\phi_\pm$ or
$z_\pm$ is advised.

\section*{Acknowledgements}

I wish to thank the Science and Technology Facilities Council (STFC, UK).

\addcontentsline{toc}{section}{References}
\bibliographystyle{format/LHCb}
\bibliography{bib/biblio}

\end{document}